%% file: zeta.tex
\documentclass[11pt]{article}

\usepackage{amsfonts}
\usepackage{amsmath} 
\usepackage{amssymb}
\usepackage[dvips]{graphics}

\newcommand{\INC}{./}

\input{\INC commands.tex}

\input{\INC numbering.tex}
\input{\INC bigpage.tex}

\renewcommand{\include}{\input}
\newcommand{\dimzeta}{\Dim_\zeta}
\newcommand{\ldimzeta}{\dim_\zeta}
\newcommand{\zetadim}{\Dim_\zeta}
\newcommand{\Zplus}{\Z^+}
\newcommand{\rep}{\mathrm{rep}}
\newcommand{\bnum}{\mathrm{bnum}}
\newcommand{\norm}[1]{\lVert #1 \rVert}
\renewcommand{\abs}[1]{\lvert #1 \rvert}

\begin{document}

\renewcommand{\emptyset}{\varnothing}
\include{zetafront}
\include{abstract}
\include{section1}

\include{section2}
\include{section3}

\include{section4}
\include{section5}


\begin{ack}
The third author thanks Tom Apostol and Giora Slutzki for useful discussions.
\end{ack}

\include{bib}

\normalsize
\newpage
\appendix
\include{proof3}
\include{proof4}
\include{proof5}

\end{document}

%% file: commands.tex


\newcommand{\PROOF}{\begin{proof}}

\newcommand{\QED}{\end{proof}}


\newcommand{\floor}[1]{ \left\lfloor #1 \right\rfloor }
\newcommand{\ceil}[1]{ \left\lceil #1 \right\rceil }

\newcommand{\myset}[2]{ \left\{ #1 \left| #2 \right. \right\} }

\newcommand{\liminfn}{\liminf\limits_{n\to\infty}}
\newcommand{\limsupn}{\limsup\limits_{n\to\infty}}


\newcommand{\N}{\mathbb{N}}
\newcommand{\Z}{\mathbb{Z}}

\newcommand{\R}{\mathbb{R}}


\newcommand{\dimh}{\mathrm{dim}_\mathrm{H}}

\renewcommand{\dim}{{\mathrm{dim}}}

\newcommand{\dimb}{{\dim}_{\mathrm{B}}}

\newcommand{\Dim}{{\mathrm{Dim}}}

\newcommand{\strSS}{S^\infty_{\mathrm{str}}}


\newcommand{\abs}{\mathrm{abs}}




\newcommand{\K}{{\mathrm{K}}}









\newcommand{\eH}{\mathrm{H}}


%% file: numbering.tex
\usepackage{amsthm}

\newtheorem{theorem}{Theorem}[section]
\newtheorem{corollary}[theorem]{Corollary}
\newtheorem{lemma}[theorem]{Lemma}
\newtheorem{proposition}[theorem]{Proposition}
\newtheorem*{claim}{Claim}

\newtheorem{observation}[theorem]{Observation}


\theoremstyle{definition}

\newtheorem*{definition}{Definition}

\newtheorem{example}[theorem]{Example}

\newtheorem*{example*}{Example}
\newtheorem*{examples*}{Examples}

\newtheorem*{remark}{Remark}

\newtheorem*{ack}{Acknowledgment}


\theoremstyle{remark}


\numberwithin{equation}{section}
\numberwithin{figure}{section}

%% file: bigpage.tex
\oddsidemargin  0.0in
\evensidemargin 0.0in
\textwidth      6.5in
\headheight     0.0in
\topmargin      -0.2in
\textheight 	8.7in

%% file: zetafront.tex
\title{ {\bf
Zeta-Dimension}
}
\author{
David Doty\footnote{Department of Computer Science,
Iowa State University,
Ames, IA 50011 USA. ddoty@cs.iastate.edu.}
\and Xiaoyang Gu\footnote{Department of Computer Science,
Iowa State University,
Ames, IA 50011 USA. xiaoyang@cs.iastate.edu. This research was supported in part by National Science Foundation Grant 0344187.}
\and Jack H. Lutz\footnote{Corresponding author. Department of Computer Science,
Iowa State University,
Ames, IA 50011 USA. lutz@cs.iastate.edu. This research was supported in part by National Science Foundation Grant 0344187.}
\and Elvira Mayordomo\footnote{Departamento de Inform\'atica e Ingenier\'ia de Sistemas,
Mar\'ia de Luna 1,
Universidad de Zaragoza,
50018 Zaragoza, SPAIN. elvira@unizar.es. This research was supported
in part by Spanish Government MEC project TIC 2002-04019-C03-03.}
\and Philippe Moser\footnote{Department of Computer Science,
Iowa State University,
Ames, IA 50011 USA. moser@cs.iastate.edu. This research was supported in part by Swiss National Science Foundation Grant PBGE2--104820.}
}
\date{}

\maketitle

%% file: abstract.tex
\begin{abstract}
The {\em zeta-dimension} of a set $A$ of positive
integers is
\[\dimzeta(A)=\inf\{s \mid \zeta_A(s)<\infty\},\]
where
\[\zeta_A(s)=\sum_{n\in A}n^{-s}.\]
Zeta-dimension serves as a fractal dimension on $\Z^+$
that extends naturally and usefully to discrete
lattices such as $\Z^d$, where $d$ is a positive integer.

This paper reviews the origins of zeta-dimension
(which date to the eighteenth and nineteenth centuries) and develops
its basic theory, with particular attention to its
relationship with algorithmic information theory. New
results presented include extended connections between
zeta-dimension and classical fractal dimensions, a 
gale characterization of zeta-dimension, and
a theorem on the zeta-dimensions of
pointwise sums and products of sets of positive
integers.

\end{abstract}

%% file: section1.tex
\begin{section}{Introduction}\label{sec:1}

Natural and engineered complex systems often
produce structures with fractal properties.
These structures may be explicitly observable
(e.g., shapes of neurons or patterns created
by cellular automata), or they may be
implicit in the behaviors of the systems
(e.g., strange attractors of dynamical systems,
Brownian trajectories in financial data, or
Boolean circuit complexity classes). In either
case, the choice of appropriate mathematical
models is crucial to understanding the
systems.

Many, perhaps most, fractal structures
are best modeled by classical fractal
geometry \cite{Falc03}, which provides top-down
specifications of many useful fractals in
Euclidean spaces and other manifolds that
support continuous mathematical methods and
attendant methods of numerical approximation.
Classical fractal geometry also includes
powerful quantitative tools, the most notable
of which are the {\em fractal dimensions}
(especially Hausdorff dimension \cite{Haus19,Falc03}
and packing dimension \cite{Tricot82,Sull84,Falc03}).
Theoretical computer scientists have recently
developed {\em effective} fractal dimensions
\cite{Lutz:EFD,Lutz:DCC,Lutz:DISS,Dai:FSD,Athreya:ESDAICC} that work
in complexity classes and other countable
settings, but these, too, are best regarded
as continuous, albeit effective, mathematical
methods.

Some fractal structures are inherently
discrete and best modeled that way. To
some extent this is already true for
structures created by cellular automata.
For the nascent theory of nanostructure
self-assembly \cite{Adle00,RotWin00}, the
case is even more compelling. This
theory models the {\em bottom-up} self-assembly
of molecular structures. The tile
assembly models that achieve this cannot
be regarded as discrete approximations
of continuous phenomena (as cellular automata
often are), because their bottom-level
units (tiles) correspond directly to discrete
objects (molecules). Fractal structures assembled
by such a model are best analyzed using
discrete tools.

This paper concerns a discrete fractal
dimension, called {\em zeta}-{\em dimension}, that works
in discrete lattices such as $\Z^d$, where
$d$ is a positive integer. Curiously, although
our work is motivated by twenty-first
century concerns in theoretical computer
science, zeta-dimension has its mathematical
origins in eighteenth and nineteenth century
number theory. Specifically, zeta-dimension
is defined in terms of a generalization of
Euler's 1737 {\em zeta}-{\em function} \cite{Euler1737} $\zeta(s)=\sum_{n=1}^\infty n^{-s}$,
defined for nonnegative real $s$
(and extended in 1859 to complex $s$ by
Riemann \cite{Riem1859}, after whom the zeta-function
is now named). Moreover, this
generalization can be formulated in terms
of Dirichlet series \cite{Diri37}, which were
developed in 1837, and one of the most
important properties of zeta-dimension
(in modern terms, the entropy characterization)
was proven in these terms by Cahen \cite{Cahen94}
in 1894.

Our objectives here are twofold. First,
we present zeta-dimension and its basic
theory, citing its origins in scattered
references, but stating things in a
unified framework emphasizing zeta-dimension's
role as a discrete fractal dimension in
theoretical computer science. Second,
we present several new results on zeta-dimension
and its interactions with classical
fractal geometry and algorithmic information
theory.

Our presentation is organized as follows.
In section \ref{sec:2}, we give an intuitive development
of zeta-dimension in the positive integers. In
section \ref{sec:3}, we extend this development in
a natural way to the integer lattices $\Z^d$,
for $d\geq 1$. In addition to reviewing known
properties of zeta-dimension, we prove discrete
analogs of two theorems of classical fractal
geometry, namely, the dimension inequalities
for Cartesian products and the total
disconnectivity of sets of dimension less
than $1$.

In section \ref{sec:4}, we discuss relationships
between zeta-dimension and classical fractal
dimensions. Many discrete fractals in $\Z^d$
have been observed to ``look like'' corresponding
fractals in $\R^d$. The most famous such
correspondence is the obvious resemblance
between Pascal's triangle modulo $2$ and
the Sierpinski triangle \cite{Stew95}. We show
how to define ``continuous versions'' of a
wide variety of self-similar discrete fractals,
and we prove that, in such cases, the
zeta-dimension of the discrete fractal is
always the Hausdorff dimension of its continuous
version. We also prove a result relating
zeta-dimension in $\Z^+$ to Hausdorff dimension
in the Baire space.

Section \ref{sec:5} concerns the relationships
between zeta-dimension and algorithmic information
theory. We review the Kolmogorov-Staiger characterization
\cite{ZvoLev70,Stai93} of the zeta-dimensions of
computably enumerable and co-computably enumerable
sets in terms of the Kolmogorov complexities (algorithmic information contents) of
their elements. We prove a theorem on the
zeta-dimensions of sets of positive integers
that are defined in terms of the digits,
or strings of digits, that can occur in
the base-$k$ expansions of their elements.
Most significantly, we prove that zeta-dimension,
like classical and effective fractal
dimensions, can be characterized in terms
of gales.
Finally, we prove a theorem on
the zeta-dimensions of pointwise
sums and products of sets of positive
integers that may have bearing on the
question of which sets of natural numbers are
definable by McKenzie-Wagner circuits \cite{McKWag03}.

Throughout this paper, $\log t=\log _2 t$, and
$\ln t=\log_e t$.

\end{section}

%% file: section2.tex
\begin{section}{Zeta-Dimension in $\Z^+$}\label{sec:2}

A set of positive integers is generally considered to
be ``small'' if the sum of the reciprocals of its
elements is finite \cite{Apos76a, HarWri79}.
Easily verified examples of such small sets
include the set of nonnegative integer powers
of $2$ and the set of perfect squares.  On the
other hand, the divergence of the harmonic series
means that the set $\Z^{+}$ of all positive integers
is not small, and a celebrated theorem of Euler
\cite{Euler1737} says that the set of all prime
numbers is not small either.

If a set is small in the above qualitative (yes/no)
sense, we are still entitled to ask, ``Exactly how
small is the set?''  This section concerns a natural,
quantitative answer to this question.  For each set
$A\subseteq\Z^{+}$ and each nonnegative real number
$s$, let
\begin{equation}\label{eq:2_1}
\zeta_{A}(s)=\sum_{n\in A}n^{-s}.
\end{equation}
Note that $\zeta_{\Z^{+}}$ is precisely $\zeta$,
the Riemann zeta-function \cite{Riem1859} (actually, Euler's original
version \cite{Euler1737} of the zeta-function, since we only consider
$\zeta_{A}(s)$ for real $s$).  The {\em zeta-dimension}
of a set $A\subseteq\Z^{+}$ is then defined to be
\begin{equation}\label{eq:2_2}
\dimzeta(A)=\inf\{ s|\zeta_{A}(s)<\infty\}.
\end{equation}
Since $\zeta_{\Z^{+}}(s)<\infty$ for all $s>1$, we have
\[ 0\leq\dimzeta(A)\leq 1 \]
for every set $A\subseteq\Zplus$.  By the results
cited in the preceding paragraph, the set of all
positive integers and the set of all prime numbers
each have zeta-dimension $1$.  Every finite set has
zeta-dimension $0$, because $\zeta_{A}(0)$ is the
cardinality of $A$.  It is easy to see that the set
of  nonnegative integer powers of $2$ also has
zeta-dimension $0$.  For a deeper example,
Wirsing's $n^{O(\frac{1}{\ln\ln n})}$ upper bound
on the number of perfect numbers not exceeding
$n$ \cite{Wirs59} implies that the set of perfect
numbers also has zeta-dimension $0$.

The zeta-dimension of a set of positive integers can
also lie strictly between $0$ and $1$.  For example,
if $A$ is the set of all perfect squares, then
$\zeta_{A}(s)=\zeta(2s)$, so $\zetadim(A)=\frac{1}{2}$.
Similarly, the set of all perfect cubes has zeta-dimension
$\frac{1}{3}$, etc.  In fact,
this argument can easily be extended to show that,
for every real number $\alpha\in[0,1]$, there exist sets
$A\subseteq\Zplus$ such that $\zetadim(A)=\alpha$.

Intuitively, we regard zeta-dimension as a fractal
dimension, analogous to Hausdorff dimension \cite{Haus19,Falc03}
or (more aptly, as we shall see) packing
dimension \cite{Tricot82,Sull84,Falc03}, on the space $\Zplus$ of
positive integers.  This intuition is supported by
the fact that zeta-dimension has the following easily
verified functional properties of a fractal dimension.
\begin{enumerate}
  \item Monotonicity: $A\subseteq B$ implies $\zetadim (A)\leq\zetadim (B)$.
  \item Stability: $\zetadim (A\cup B)=\max\{\zetadim (A), \zetadim (B)\}$.
  \item Translation invariance: For each $k\in\Zplus$, $\zetadim (k+A)=\zetadim (A)$, where     $k+A=\{k+n|n \in A\}$.
  \item Expansion invariance: For each $k\in\Zplus$, $\zetadim (kA)=\zetadim (A)$, where     $kA=\{kn|n\in A\}$.
\end{enumerate}

Equation \eqref{eq:2_1} can be written as a Dirichlet
series
\begin{equation}\label{eq:2_3}
\zeta_A(s)=\sum_{n=1}^\infty f(n) n^{-s}
\end{equation}
in which $f$ is the characteristic function of $A$.
In the terminology of analytic number theory,
\eqref{eq:2_2} then says that the zeta-dimension of
$A$ is the {\em abscissa of convergence} of the
series \eqref{eq:2_3} \cite{Knop90,HarWri79,Apos76a,Apos76b}.
In this sense, zeta-dimension was introduced
in 1837 by Dirichlet \cite{Diri37}. The following
useful characterization of zeta-dimension was
proven in this more general setting in 1894.

\begin{theorem}[entropy characterization of zeta-dimension -- Cahen \cite{Cahen94}; see also \cite{Knop10,Knop90,HarWri79,Apos76a,Apos76b}]\label{th:2_1}
For all $A\subseteq \Z^+$,
\begin{equation}\label{eq:2_4}
\dimzeta(A)=\limsupn\frac{\log|A\cap \{1, \dots, n\}|}{\log n}.
\end{equation}
\end{theorem}
\begin{example}\label{ex:2_2}
The set $C'$, consisting of all positive integers
whose ternary expansions do not contain a $1$,
can be regarded as a discrete analog of the
Cantor middle thirds set $C$, which consists of
all real numbers in $[0,1]$ who ternary
expansions do not contain a $1$. Theorem \ref{th:2_1}
implies immediately that $C'$ has
zeta-dimension $\tfrac{\log 2}{\log 3}\approx 0.6309$, which is exactly the
classical fractal (Hausdorff or packing) dimension
of $C$. We will see in section \ref{sec:4} that this
is not a coincidence, but rather a special
case of a general phenomenon.
\end{example}

By Theorem \ref{th:2_1} and routine calculus, we
have
\begin{equation}\label{eq:2_5}
\dimzeta(A)=\limsupn\frac{\log |A\cap \{1,\dots, 2^n\}|}{n}
\end{equation}
and
\begin{equation}\label{eq:2_6}
\dimzeta(A)=\limsupn\frac{\log |A\cap \{2^n,\dots, 2^{n+1}-1\}|}{n}
\end{equation}
for all $A\subseteq \Z^+$. The right-hand side of
\eqref{eq:2_6} has been called the ({\em channel}) {\em capacity}
of $A$ and the {\em entropy} ({\em rate}) of $A$ \cite{Shannon48,
Kuic70,Eile74,Conn75,deLu75,Stai93}. In
particular, Staiger \cite{Stai93} (see also \cite{Hitchcock:EFDFAA}) rediscovered \eqref{eq:2_6}
as a characterization of the entropy of $A$.

The following section shows how
to extend zeta-dimension to the integer
lattices $\Z^d$, for $d\geq 1$.

\end{section}

%% file: section3.tex
\begin{section}{Zeta-Dimension in $\Z^d$}\label{sec:3}

For each $\vec{n}=(n_1,\dots, n_d)\in \Z^d$, where $d$ is
a positive integer, let $\norm{\vec{n}}$ be the
Euclidean distance from the origin to $\vec{n}$,
i.e.,
\begin{equation}\label{eq:3_1}
\norm{\vec{n}}=\sqrt{n_1^2+\cdots + n_d^2}.
\end{equation}
For each $A\subseteq \Z^d$, define the $A$-{\em zeta}-{\em function}
$\zeta_A:[0,\infty)\rightarrow[0,\infty]$ by
\begin{equation}\label{eq:3_2}
\zeta_A(s)=\sum_{\vec{0}\neq \vec{n}\in A}\norm{\vec{n}}^{-s}
\end{equation}
for all $s\in[0,\infty)$, and define the {\em zeta}-{\em dimension}
of $A$ to be
\begin{equation}\label{eq:3_3}
\dimzeta(A)=\inf\{s\mid \zeta_A(s)<\infty\}.
\end{equation}

Note that, if $d=1$ and $A\subseteq \Z^+$, then
definitions \eqref{eq:3_2} and \eqref{eq:3_3} agree with
definitions \eqref{eq:2_1} and \eqref{eq:2_2}, respectively. The
zeta-dimension that we have defined in $\Z^d$
is thus an extension of the one that
was defined in $\Z^+$.

\begin{observation}\label{ob:3_1}
For all $d\in\Z^+$ and $A\subseteq \Z^d$,
\[0\leq \dimzeta(A)\leq d.\]
\end{observation}

We next note that zeta-dimension has
key properties of a fractal dimension in
$\Z^d$. We state the invariance property
a bit more generally than in section \ref{sec:2}.

\begin{definition}
A function $f:\Z^d\rightarrow \Z^d$ is
{\em bi}-{\em Lipschitz} if there exists $\alpha,\beta\in(0,\infty)$
such that, for all $\vec{m}$, $\vec{n}\in \Z^d$,
\[\alpha\norm{\vec{m}-\vec{n}}\leq 
\norm{f(\vec{m})-f(\vec{n})}\leq \beta \norm{\vec{m}-\vec{n}}.\]
\end{definition}

\begin{observation}[fractal properties of zeta-dimension]\label{ob:3_2}
Let $A, B\subseteq \Z^d$.
\begin{enumerate}
\item
Monotonicity: $A\subseteq B$ implies $\dimzeta(A)\leq \dimzeta(B)$.
\item
Stability: $\dimzeta(A\cup B)=\max \{ \dimzeta(A), \dimzeta(B)\}$.
\item
Lipschitz invariance: If $f:\Z^d\rightarrow \Z^d$ is bi-Lipschitz,
then $\dimzeta(f(A))=\dimzeta(A)$.
\end{enumerate}
\end{observation}

For $A\subseteq \Z^d$ and $I\subseteq [0,\infty)$, let
\[A_I=\{\vec{n}\in A\mid \norm{\vec{n}}\in I\}.\]
Then the
Dirichlet series
\begin{equation}\label{eq:3_4}
\zeta^D_A(s)=\sum_{n=1}^\infty \abs{A_{[n,n+1)}}n^{-s}=\sum_{\vec{0}\neq\vec{n}\in A}\floor{\norm{\vec{n}}}^{-s},
\end{equation}
converges exactly when $\zeta_A(s)$ converges,
so equation \eqref{eq:3_3} says that $\dimzeta(A)$ is the
abscissa of convergence of this series. Cahen's
1894 characterization of this abscissa thus
gives us the following extension of Theorem \ref{th:2_1}.

\begin{theorem}[entropy characterization of zeta-dimension in $\Z^d$ -- Cahen \cite{Cahen94}]\label{th:3_3}
For all $A\subseteq \Z^d$,
\begin{equation}\label{eq:3_5}
\dimzeta(A)=\limsupn \frac{\log \abs{A_{[1,n]}}}{\log n}.
\end{equation}
\end{theorem}

As in $\Z^+$, it follows immediately by
routine calculus that
\begin{equation}\label{eq:3_6}
\dimzeta(A)=\limsupn\frac{\log\abs{A_{[1,2^n]}}}{n}
\end{equation}
and
\begin{equation}\label{eq:3_7}
\dimzeta(A)=\limsupn\frac{\log\abs{A_{[2^n,2^{n+1})}}}{n}
\end{equation}
for all $A\subseteq \Z^d$.
Willson \cite{Will84} has used (a quantity formally
identical to) the right-hand side of \eqref{eq:3_6} as
a measure of the {\em growth}-{\em rate} {\em dimension} of a
cellular automaton.


We next note that ``subspaces'' of $\Z^d$
have the ``correct'' zeta-dimensions.

\begin{theorem}\label{th:3_w}
If $\vec{m_1}, \dots, \vec{m_k}\in\Z^d$ are linearly
independent (as vectors in $\R^d$) and
\[S=\{ a_1\vec{m_1}+\cdots +a_k\vec{m_k} \mid a_1,\dots, a_k\in\Z  \},\]
then $\dimzeta(S)=k$.
\end{theorem}

By translation invariance, it follows
that ``hyperplanes'' in $\Z^d$ also have the
``correct'' zeta-dimensions.

The Euclidean norm \eqref{eq:3_1} is sometimes
inconvenient for calculations. When desirable,
the $L^1$ norm,
\[\norm{\vec{n}}_1=\abs{n_1}+\cdots +\abs{n_d},\]
can be used in its place. That is, if we
define the $L^1$ $A$-{\em zeta}-{\em function} $\zeta_A^{L^1}$ by
\[\zeta_A^{L^1}(s)=\sum_{\vec{0}\neq \vec{n}\in A} \norm{\vec{n}}_1^{-s},\]
then
\[2^{-s}\zeta_A(s)\leq \zeta_A^{L^1}(s)\leq \zeta_A(s)\]
holds for all $s\in[0,\infty)$, so
\[\dimzeta(A)=\inf\{ s\mid \zeta_A^{L^1}(s)<\infty\}.\]
The entropy characterizations \eqref{eq:3_5}, \eqref{eq:3_6}, and
\eqref{eq:3_7} also hold with each set $A_I$ replaced
by the set
\[A_I^{L^1}=\{\vec{n}\in A\mid \norm{\vec{n}}_1\in I\}.\]

\begin{example}[Pascal's triangle modulo $2$]\label{ex:3_4}
Let 
\[A=\{ (m,n)\in \N^2\mid \tbinom{m+n}{m}\equiv 1 \mod 2\}.\]
Then it is easy to see that $\abs{A_{[1,2^n]}^{L^1}}=3^n$
for all $n\in\N$, whence the $L^1$ version of
\eqref{eq:3_6} tells us that $\dimzeta(A)=\log 3\approx 1.5850$. This
is exactly the fractal (Hausdorff or packing)
dimension of the Sierpinski triangle that $A$
so famously resembles \cite{Stew95}. This connection
will be further illuminated in section \ref{sec:4}.
\end{example}

In order to examine the zeta-dimensions of
Cartesian products, we define the {\em lower zeta}-{\em dimension}
of a set $A\subseteq \Z^+$ to be
\begin{equation}\label{eq:3_8}
\ldimzeta(A)=\liminfn \frac{\log\abs{A_{[1,n]}}}{\log n}.
\end{equation}
By Theorem \ref{th:3_3}, $\ldimzeta(A)$ is a sort of
dual of $\dimzeta(A)$.
By routine
calculus, we also have
\begin{equation}\label{eq:3_9}
\ldimzeta(A)=\liminfn\frac{\log\abs{A_{[1,2^n]}}}{n},
\end{equation}
i.e., the dual of equation \eqref{eq:3_6} holds. Note,
however, that the dual of equation \eqref{eq:3_7} does
{\em not} hold in general.

The following theorem is exactly analogous to
a classical theorem on the Hausdorff and packing dimensions
of Cartesian products \cite{Falc03}.
\begin{theorem}\label{th:3_5}
For all $A\subseteq \Z^{d_1}$ and $B\subseteq \Z^{d_2}$,
\begin{align*}
\ldimzeta(A)+\ldimzeta(B)&\leq \ldimzeta(A\times B)\\
&\leq \ldimzeta(A)+\dimzeta(B)\\
&\leq \dimzeta(A\times B)\\
&\leq \dimzeta(A)+\dimzeta(B).
\end{align*}
\end{theorem}

Although connectivity properties play an important
role in classical fractal geometry, their role in
discrete settings like $\Z^d$ will perforce be
more limited. Nevertheless, we have the following.
Given $d, r\in\Z^+$, and points $\vec{m}, \vec{n}\in \Z^d$,
an $r$-{\em path} from $\vec{m}$ to $\vec{n}$ is a sequence
$\pi=(\vec{p_0},\dots, \vec{p_l})$ of points $\vec{p_i}\in\Z^d$ such
that $\vec{p_0}=\vec{m}$, $\vec{p_l}=\vec{n}$,
and $\norm{\vec{p_i}-\vec{p_{i+1}}}\leq r$
for all $0\leq i<l$. Call a set  $A\subseteq \Z^d$
{\em boundedly connected} if there exists $r\in\Z^+$
such that, for all $\vec{m},\vec{n}\in A$, there
is an $r$-path $\pi=(\vec{p_0},\dots, \vec{p_l})$ from $\vec{m}$
to $\vec{n}$ in which $\vec{p_i}\in A$ for all $0\leq i\leq l$.

A result of classical fractal geometry
says that any set of dimension less than
$1$ is totally disconnected. The following
theorem is an analog of this for zeta-dimension.

\begin{theorem}\label{th:3_6}
Let $d\in\Z^+$ and $A\subseteq \Z^d$. If
$\dimzeta(A)<1$, then no infinite subset of $A$
is boundedly connected.
\end{theorem}
The next section examines the relationships
between zeta-dimension and classical fractal
dimensions in greater detail.

\end{section}

%% file: section4.tex
\section{Zeta-Dimension and Classical Fractal Dimension}\label{sec:4}

The following result shows that the agreement between zeta-dimension
and Hausdorff dimension
noticed in  Examples \ref{ex:2_2} and \ref{ex:3_4} are
instances of a more general phenomenon:
Given any discrete fractal with enough self similarity, its zeta-dimension
is equal to the Hausdorff dimension of its classical version.
Previous results along these lines were proven by Willson
\cite{Will84,Will87}, for the special case of sets that are obtained from additive cellular automata.

The following states what is meant by self-similarity precisely.

\begin{definition} \label{def:self-similar}
Let $c,d\in\N$, $F\subset \N^d$. $F$ is a $c$-discrete self similar fractal, if there exists a function
	$$S:\{ 1,2,\cdots ,c \}^d \rightarrow \{ \text{no}, R_0, R_1, R_2, R_3\}$$ such that
	$S(1, 1, \cdots , 1)=R_0$, and for every integer  $k$ and every
	$(i_1, \cdots, i_d)\in \{ 1,2,\cdots ,c \}^d$,
	$$F\cap  C^k_{i_1,i_2,\cdots,i_d } =
	\begin{cases}
		R_j(C^k_{1, \cdots ,1}) & \text{ if $S(i_1, \cdots, i_d)=R_j$},\\
		\varnothing & \text{ if $S(i_1, \cdots, i_d)=$ no}
	\end{cases}
	$$
	where $R_j$ ($j=0, \cdots , 3$) is a rotation of angle $j\pi /2$, and
	$$C^k_{i_1,i_2,\cdots,i_d }=[(i_1-1)c^k+1,i_1c^k] \times \cdots \times [(i_d-1)c^k+1,i_dc^k]$$
	is a $d$-dimensional cube of side $c$.
\end{definition}

	There are many ways to generalize the above definition 
	including statistical similarity, multiple patterns, fractal curves constructed from a generator \cite{Falc03},
	multiple contraction ratio (of the form $c_1, \cdots , c_n $ where $c_i|c_n$ for $i<n$). Also the preserved
	cube does not need to be $C^k_{1, \cdots ,1}$, but can be any cube $C$, in which case the discrete
	fractal will grow in $\mathbb{Z}^d$ starting from $C$. It is easy to see that the following result still holds for
	those more general cases.

	Given any $c$-discrete self similar fractal $F\subset \N^d$, we construct its continuous analogue
	$\mathbb{F}\subset [0,1]^d$ recursively, via the following contraction $T:x\mapsto \frac{1}{c}x$.
	$\mathbb{F}_0=[0,1]$ and $\mathbb{F}_k=T^{(k)}(F\cap [1,c^k]^d)$, where $T^{(k)}=T\circ \cdots \circ T$, denotes $k$
	iterations of $T$. The fractal $\mathbb{F}=\lim_{k\rightarrow\infty}\mathbb{F}_k$ obtained by this construction is a self-similar
	continuous fractal with contraction ratio $1/c$. The following result shows that the zeta-dimension of the discrete fractal
	is equal to the Hausdorff dimension of the continuous one.

\begin{theorem} \label{th:4_1}
	If $c, d,F, \mathbb{F}$ are as above, then
	$$\dimzeta(F)=\dimh(\mathbb{F}).$$
\end{theorem}

\newcommand{\real}{\mathrm{real}}

	The following result gives a relationship between zeta-dimension and dimension in the Baire space.
	We consider the Baire space $\N^{\infty}$ representing total functions from $\N$ to $\N$
	in the obvious way. Given $w\in \N^*$, let
	$C_w= \{ z\in\N^{\infty}|w\sqsubset z \} $. We define
	$\real :\N^{\infty} \rightarrow [0,1]$ by 
	$$\real (z)=\cfrac{1}{(z_0+1)+\cfrac{1}{(z_1+1)+\dotsb}}.$$
	The cylinder generated by $w$ is the interval
	$\Delta (w)= \{ x\in [0,1]| x = \real (z), w \sqsubset z \}$.

	A subprobability supermeasure on $\N^{\infty}$ is a function
	$p: \N^* \rightarrow [0,1]$ such that $p(\lambda)\leq 1$ and for each $w \in \N^*$,
	$p(w)\geq \sum_{n}p(wn)$.

	For each subprobability supermeasure $p$ we can define a Hausdorff dimension and a packing dimension
on $\N^{\infty}$,
$\dim_{p}$ and $\Dim_p$, using the metric $\rho$ defined as $\rho (z, z')= p(w)$ for $w\in\N^*$ the longest
common prefix of $z,z'\in\N^{\infty}$.

	\emph{Gauss measure} is defined on each $E\subseteq \R$ as
	$$\gamma (E)= \frac{1}{\ln 2}\int_{E}\frac{dt}{1+t}.$$
	We will abuse notation and use $\gamma (w) = \gamma (\real (C_w))$ for each $w\in\N^*$.
	Notice that $\gamma (\lambda) =1$ and therefore $\gamma$ is a probability measure on $\N^{\infty}$.

	\begin{remark}
				Let $\mu$ denote the Lebesgue measure on $\R$, and let $\mu(w)=\mu(\real (C_w))$
				for each $w\in \N^*$, then
				$$\frac{\mu (w)}{2\ln 2}\leq \gamma (w) \leq \frac{\mu (w)}{\ln 2},$$
so $\mu$ and $\gamma$ give equivalent Hausdorff dimensions.
	\end{remark}

	Define $F_A=\{ f:\N\rightarrow\N | f(\N)\subseteq A \text{ and }\lim_{n\rightarrow\infty}f(n)=\infty\}$,
	for each $A\subseteq \Z^{+}$.
	The following result relates zeta-dimension to Gauss-dimension.
\begin{theorem} \label{th:4_2}
$\dimzeta(A)=2 \cdot \dim_{\gamma} (F_A)=2\cdot\Dim_\gamma (F_A)$.
\end{theorem}

%% file: section5.tex
\begin{section}{Zeta-Dimension and Algorithmic Information}\label{sec:5}

The entropy characterization of zeta-dimension
(Theorem \ref{th:3_3}) already indicates a strong connection
between zeta-dimension and information theory.
Here we explore further such connections. The first concerns
the zeta-dimensions of sets of positive integers that are
defined in terms of the digits, or strings of digits, that
can appear in the base-$k$ expansions of their elements.
We write $\mathrm{rep}_{k} (n)$ for the base-$k$ expansion
$(k \geq 2)$ of a positive integer $n$.  Conversely, given
a nonempty string $w\in\{0,1,\cdots,k-1\}^{*}$ that does
not begin with $0$, we write $\mathrm{num}_{k}(w)$ for the
positive integer whose base-$k$ expansion is $w$.

A {\em prefix set} over an alphabet $\Sigma$ is a set
$B\subseteq\Sigma^{*}$ such that no element of $B$ is
a proper prefix of another element of $B$.  An
{\em instantaneous code} is a nonempty prefix set that
does not contain the empty string.

\begin{theorem}\label{th:5_1}
Let $\Sigma = \{0,1, \cdots ,k-1\}$, where $k\geq 2$.
Assume that $\varnothing \neq \Delta \subseteq \Sigma - \{0\}$
and that $B \subseteq \Sigma^{*}$ is a finite instantaneous
code, and let
\[ A=\{n\in\Zplus | \mathrm{rep}_{k}(n) \in \Delta B^{*} \}. \]
Then
\[ \zetadim (A) = s^{*}, \]
where $s^{*}$ is the unique solution of the equation
\[ \sum_{w \in B} k^{-s^{*}\abs{w}} = 1. \]
\end{theorem}

\begin{corollary}\label{co:5_2}
Let $\Sigma = \{0,1, \cdots ,k-1\}$, where $k\geq 2$.
If $\Gamma \subseteq \Sigma$ and $\Gamma \not\subseteq \{0\}$ and
\[ A = \{n \in \Zplus | \mathrm{rep}_{k}(n) \in \Gamma^{*} \}, \]
then
\[ \zetadim (A) = \frac{\ln\abs{\Gamma}}{\ln k} . \]
\end{corollary}

\begin{example}\label{ex:5_3}
Corollary \ref{co:5_2} gives a quantitative articulation of the
``paradox of the missing digit''\cite{HarWri79}.  If $A$
is the set of positive integers in whose decimal expansions
some particular digit, such as $7$, is missing, then a naive
intuition might suggest that $A$ contains ``most'' integers,
but $A$ has long been known to be small in the sense that the
sum of the reciprocals of its elements is finite (i.e.,
$\zeta_{A} (1) < \infty$).  In fact, Corollary \ref{co:5_2} says
that $\zetadim (A) = \frac{\ln 9}{\ln 10} \approx 0.9542$,
a quantity somewhat smaller than, say, the zeta-dimension
of the set of prime numbers.
\end{example}

The main connection between zeta-dimension
and {\em algorithmic} information theory is a theorem
of Staiger \cite{Stai93} relating entropy to Kolmogorov
complexity. To state Staiger's theorem in our
present framework, we define the {\em Kolmogorov complexity}
$\K(\vec{n})$ of a point $\vec{n}\in\Z^d$ to be
the length of a shortest program $\pi\in\{0,1\}^*$ such
that, when a fixed universal self-delimiting
Turing machine $U$ is run with $(\pi, d)$ as its
input, $U$ outputs $\vec{n}$ (actually, some straightforward
encoding of $\vec{n}$ as a binary string)
and halts after finitely many computation steps.  Detailed
discussions of Kolmogorov complexity's definition, fundamental
properties, history, significance, and applications appear in
the definitive textbook by Li and Vitanyi \cite{LiVi97}.  As
we have already noted, $\K(\vec{n})$ is a measure of the
{\em algorithmic information content} of $\vec{n}$.

For $\vec{0}\neq \vec{n}\in \Z^d$, we write $l(\norm{\vec{n}})$ for the
length of the standard binary expansion (no leading
zeroes) of the positive integer $\floor{\norm{\vec{n}}}$.

If $f:\Z^d\rightarrow[0,\infty)$ and $A\subseteq \Z^d$, then the
{\em limit superior} of $f$ on $A$ is
\[\limsup_{\vec{n}\in A}f(\vec{n})=\lim_{k\rightarrow \infty}\sup f(A_{[k,\infty]}).\]
Note that this is $0$ if $A$ is finite.

\begin{theorem}[Kolmogorov \cite{ZvoLev70}, Staiger \cite{Stai93}]\label{th:5_4}
For every $A\subseteq \Z^d$,
\[\dimzeta(A)\leq \limsup_{\vec{n}\in A}\frac{\K(\vec{n})}{l(\norm{\vec{n}})},\]
with equality if $A$ or its complement is computably enumerable.
\end{theorem}

In the case where $d=1$ and $A\subseteq \Z^+$, Theorem
\ref{th:5_4} says that, if $A$ is $\Sigma^0_1$ or $\Pi^0_1$, then
\[\dimzeta(A)=\limsup_{n\in A}\frac{\K(n)}{l(n)},\]
where $l(n)$ is the length of the binary representation
of $A$. Kolmogorov \cite{ZvoLev70} proved this for $\Sigma^0_1$ sets,
and Staiger \cite{Stai93} proved it for $\Pi^0_1$ sets. The
extension to $A\subseteq \Z^d$ for arbitrary $d\in\Z^+$ is
routine.

As Staiger has noted, Theorem \ref{th:5_4}
cannot be extended to $\Delta^0_2$ sets,
because an oracle for the halting problem can
easily be used to decide a set $B\subseteq \Z^+$
such that, for each $k\in\Z^+$, $B_{[2^k, 2^{k+1}]}$ contains
exactly one integer $n$, and this $n$ also
satisfies $\K(n)\geq k$. Such a set $B$ is a
$\Delta^0_2$ set satisfying $\dimzeta(B)=0<1=\limsup_{n\in B}\frac{\K(n)}{l(n)}$.


Classical Hausdorff and packing dimensions
were recently characterized in terms of gales,
which are betting strategies with a parameter $s$
that quantifies how favorable the payoffs are
\cite{Lutz:DCC,Athreya:ESDAICC}. These characterizations have
played a central role in many recent
studies of effective fractal dimensions in
algorithmic information theory and computational
complexity theory \cite{Lutz:EFD}. We show here that
zeta-dimension also admits such a
characterization.

Briefly, given $s\in[0,\infty)$, an $s$-{\em gale} is
a function $d:\{0,1\}^*\rightarrow [0,\infty)$ satisfying
$d(w)=2^{-s}[d(w0)+d(w1)]$ for all $w\in \{0,1\}^*$.
For purposes of this paper, an $s$-gale $d$
{\em succeeds} on a positive integer $n$ if
$d(w)\geq 1$, where $w$ is the standard binary representation
of $n$.

\begin{theorem}[gale characterization of zeta-dimension]\label{th:5_5}
For all $A\subseteq \Z^+$,
\[\dimzeta(A)=\inf \{ s\mid \text{there is an $s$-gale $d$ that succeeds on every element of $A$}\}.\]
\end{theorem}

Our last result is a
theorem on the zeta-dimensions of pointwise sums
and products of sets of positive integers. For
$A,B\subseteq \Z^+$, we use the notations
\begin{align*}
&A+B=\{a+b\mid a\in A \text{ and } b\in B\},\\
&A*B=\{ ab\mid a\in A\text{ and } b\in B\}.
\end{align*}
The first equality in the following theorem is due to
Staiger \cite{Stai93}.

\begin{theorem}\label{th:5_6}
If $A, B\subseteq \Z^+$ are nonempty, then
\[\dimzeta(A*B)=\max
\{\dimzeta(A),\dimzeta(B)\}\leq \dimzeta(A+B)\leq \dimzeta(A)+\dimzeta(B),
\]
and the inequalities are tight in the strong sense that,
for all $\alpha, \beta, \gamma\in[0,1]$ with 
$\max\{ \alpha,\beta\}\leq \gamma\leq \alpha+\beta$,
there exist $A,B\subseteq \Z^+$ with $\dimzeta(A)=\alpha$,
$\dimzeta(B)=\beta$, and $\dimzeta(A+B)=\gamma$.
\end{theorem}

We close with a question concerning
circuit definability of sets of natural numbers, a notion introduced
recently by McKenzie and Wagner \cite{McKWag03}.  Briefly, a
McKenzie-Wagner {\em circuit} is a combinational circuit
(finite directed acyclic graph) in which the inputs are singleton sets
of natural numbers, and each gate is of one of five types.  Gates of
type $\cup$, $\cap$, $+$, and $*$ have indegree $2$ and compute set
union, set intersection, pointwise sum,
and pointwise product, respectively.
Gates of type $^{-}$  have indegree $1$ and compute set complement.
Each such circuit {\em defines} the set of natural
numbers computed at its designated output gate in the obvious way.
The fact that $0$ is a natural number is crucial in this model.
Interesting sets that are known to be definable
in this model include the set of primes, the set of powers of a given
prime, and the set of counterexamples to Goldbach's conjecture.  Is
there a zero-one law, according to which every set definable by a
McKenzie-Wagner circuit has zeta-dimension $0$ or $1$?  Such a law
would explain the fact that the set of perfect squares is not known
to be definable by such circuits.
Theorem \ref{th:5_6} suggests that a zero-one law, if true, will not be proven
by a trivial induction on circuits.

\end{section}

%% file: bib.tex
\bibliographystyle{abbrv}
\small
\input{\INC allbibs.tex}

%% file: allbibs.tex
\newcommand{\lINC}{\INC}
\bibliography{\lINC dim,\lINC rbm,\lINC main,\lINC random,\lINC dimrelated}

%% file: proof3.tex
\section{Appendix -- Zeta-Dimension in $\Z^d$}

\begin{proof}[Proof of Theorem \ref{th:3_w}]
Assume the hypothesis.
By standard results in the geometry of numbers,
there exist constants $\alpha, \beta\in(0,\infty)$ such that,
for all $n\in\Z^+$,
\[\alpha n^k\leq \abs{S_{[1,n]}}\leq \beta n^k.\]
It follows by Theorem \ref{th:3_3} that $\dimzeta(S)=k$.
\end{proof}

\begin{proof}[Proof of Theorem \ref{th:3_5}]
	The following is easy to show.
	\begin{claim}
		Let $A\subseteq \Z^{d_1}$, $B\subseteq \Z^{d_2}$ and $n\in\N$, then
		$$A_{[1,n]}\times B_{[1,n]} \subseteq (A\times B)_{[1,2n]} \subseteq A_{[1,2n]}\times B_{[1,2n]}$$
		i.e.
		$$|A_{[1,n]}|\cdot |B_{[1,n]}| \leq |(A\times B)_{[1,2n]}| \leq |A_{[1,2n]}|\cdot |B_{[1,2n]}|.$$
	\end{claim}

	Let us prove the first inequality.
	\begin{align*}
		\ldimzeta (A) + \ldimzeta (B) &=
		\liminf_{n\rightarrow\infty} \frac{\log |A_{[1,n]}|}{\log n} +
		\liminf_{n\rightarrow\infty} \frac{\log |A_{[1,n]}|}{\log n} \\
		& \leq \liminfn \frac{\log (|A_{[1,n]}| \cdot |B_{[1,n]}|)}{\log n}\\
		& \leq \liminfn \frac{\log (|(A\times B)_{[1,2n]}|)}{\log n}\\
		& \leq \liminfn \frac{\log (|(A\times B)_{[1,2n]}|)}{\log 2n}
		= \ldimzeta (A\times B)
	\end{align*}

	For the second inequality we have
	\begin{align*}
		\ldimzeta (A\times B)  &=
		\liminf_{n\rightarrow\infty} \frac{\log |(A\times B)_{[1,n]}|}{\log n} \\
		&\leq \liminf_{n\rightarrow\infty} \frac{\log (|A_{[1,n]}|\cdot |B_{[1,n]}|)}{\log n} \\
		&= \liminf_{n\rightarrow\infty} \frac{\log |A_{[1,n]}|+ \log |B_{[1,n]}|}{\log n} \\
		&\leq \liminf_{n\rightarrow\infty} \frac{\log |A_{[1,n]}|}{\log n}
		+  \limsup_{n\rightarrow\infty} \frac{\log |B_{[1,n]}|}{\log n}
		= \ldimzeta (A)+ \dimzeta (B)
	\end{align*}

	For the third inequality we have
	\begin{align*}
		\ldimzeta (A)+ \dimzeta (B) &=
		\liminfn \frac{\log |A_{[1,n]}|}{\log n}
		+  \limsupn \frac{\log |B_{[1,n]}|}{\log n}\\
		& \leq \limsupn \frac{\log |A_{[1,n]}| +\log |B_{[1,n]}|}{\log n}\\
		& = \limsupn \frac{\log (|A_{[1,n]}|\cdot  |B_{[1,n]}|)}{\log n}\\
		& \leq \limsupn \frac{\log (|(A\times B)_{[1,2n]}|)}{\log n}\\
		& \leq \limsupn \frac{\log (|(A\times B)_{[1,2n]}|)}{\log 2n}
		= \dimzeta (A\times B)
	\end{align*}
	For the last inequality we have
	\begin{align*}
		\ldimzeta (A\times B)
		&= \limsupn \frac{\log (|(A\times B)_{[1,n]}|)}{\log n}\\
		&\leq \limsupn \frac{\log |A_{[1,n]}|}{\log n}+\frac{\log |B_{[1,n]}|}{\log n} \\
		&\leq \limsupn \frac{\log |A_{[1,n]}|}{\log n}
		+\limsupn\frac{\log |B_{[1,n]}|}{\log n}
		= \dimzeta (A)+ \dimzeta (B)
	\end{align*}
\end{proof}

\begin{proof}[Proof of Theorem \ref{th:3_6}]
Let $A\subseteq \Z^d$, and let
$C$ be an infinite, boundedly connected subset of
$A$. It suffices to prove that $\dimzeta(A)\geq 1$.

Write $C=\{ \vec{n_k} \mid k\in\N\}$. Since $C$ is
boundedly connected, there is, for each
$k\in\N$, an $r$-path $\pi_k$ from $\vec{n_k}$ to $\vec{n_{k+1}}$,
all of whose points are in $C$. Inserting
those paths into the list $\vec{n_0}, \vec{n_1}, \dots$,
we get an expanded  list $\vec{m_0}, \vec{m_1},\dots$ of
points in $C$ such that (i) every point of
$C$ appears in the list $\vec{m_0}, \vec{m_1}, \dots$; and (ii)
for all $k\in\N$, $\norm{\vec{m_k},\vec{m_{k+1}}}\leq r$.
If we now delete from
the list $\vec{m_0}, \vec{m_1}, \dots$ each $\vec{m_k}$ that
has appeared earlier in the list, then we
obtain an enumeration $\vec{p_0},\vec{p_1},\dots $ of
$C$ in which there is no repetition and
\[\norm{\vec{p_k}}\leq \norm{\vec{p_0}} +kr\]
holds for all $k\in\N$. It follows that
\begin{align*}
\zeta_A(1)&\geq \zeta_C(1)\\
&=\sum_{k=0}^\infty \norm{\vec{p_k}}^{-1}\\
&\geq \sum_{k=0}^\infty \frac{1}{\norm{\vec{p_0}}+kr}\\
&=\infty,
\end{align*}
whence $\dimzeta(A)\geq 1$.
\end{proof}

%% file: proof4.tex
\section{Appendix -- Zeta-Dimension and Classical Fractal Dimension}
\begin{proof}[Proof of Theorem \ref{th:4_1}]
	Consider $F_k=[1,c^k]^d$ and let
	$$B(F_k)= \frac{\log |F_k|}{k \log c} \ \text{ and } \ B(F)= \lim_{k\rightarrow\infty}B(F_k).$$
	\begin{claim}
		$|F_k| = |S^{-1}(\{ R_0, \cdots , R_3 \} )|^k$.
	\end{claim}
	We prove the claim by induction. The claim is true for $k=1$; let $k\in \N$, we have
	$$|F_k|= |F_{k-1}|\cdot |S^{-1}(\{ R_0, \cdots , R_3 \} )|= |S^{-1}(\{ R_0, \cdots , R_3 \} )|^{k}.$$
	This proves the claim.

	Let $Y=|S^{-1}(\{ R_0, \cdots , R_3 \} )|$. By the claim,
	$$B(F)= \lim_{k\rightarrow\infty}B(F_k)= \lim_{k\rightarrow\infty} \frac{\log |F_k|}{k \log c}= \frac{\log Y}{\log c}.$$
	\begin{claim}
		$\dimzeta(F)= B(F)$.
	\end{claim}
	To prove the claim consider $D_k=F_{k+1} - F_{k}$. We have $|D_{k}|= Y^k(Y-1)$. For a tuple
	$(m_1, \cdots , m_d) \in D_k$ we have
	$$ dc^k \leq m_1+ \cdots + m_d \leq dc^{k+1}$$ thus
	$$ d^{-s}c^{-s(k+1)} \leq (m_1+ \cdots + m_d)^{-s} \leq d^{-s}c^{-sk}$$ i.e.
	$$ |D_k| d^{-s}c^{-s(k+1)} \leq \zeta_{D_k}(s) \leq |D_k| d^{-s}c^{-sk}$$ therefore
	$$ Y^k(Y-1)d^{-s}c^{-s(k+1)} \leq \zeta_{D_k}(s) \leq  Y^k(Y-1) d^{-s}c^{-sk}$$ thus
	$$ a\sum_{k\geq 1}(Yc^{-s}) \leq \dimzeta(F) \leq b\sum_{k\geq 1}(Yc^{-s})$$
	where $a, b$ are constants. The convergence radius  of the upper sum gives the zeta-dimension of $F$, i.e.
	is solution of the equation $Yc^{-s}=1$, thus $s=\log Y /\log c$, which proves the claim.

	\begin{claim}
		$\dimh(\mathbb{F})= B(F)$.
	\end{claim}
	The box dimension of $\mathbb{F}$ is given by
	$$\dimb(\mathbb{F})=\lim_{k\rightarrow \infty}\frac{\log N_{c^{-k}}(\mathbb{F})}{k\log c}$$
	where $N_{c^{-k}}$ is the number of $d$-mesh cubes of side $c^{-k}$ of the form
	$$M^{k}_{m_1, \cdots , m_d}=[m_1c^{-k},(m_1+1)c^{-k}]\times \cdots [m_dc^{-k},(m_d+1)c^{-k}], \ \text{ where } \ m_i\in\N$$
	required to cover $\mathbb{F}$.

	Since $\mathbb{F}\subset \mathbb{F}_k$ we have $N_{c^{-k}}(\mathbb{F})\leq N_{c^{-k}}(\mathbb{F}_k)$.
	Moreover the number of mesh cubes $M^{k}_{m_1, \cdots , m_d}$ required to cover $\mathbb{F}_k$ is equal to
	the number required to cover $\mathbb{F}_{k+j}$ for any integer $j$, because
	$\mathbb{F}_k\cap M^{k}_{m_1, \cdots , m_d} \neq\emptyset$ implies $\mathbb{F}_{k+j} \cap M^{k}_{m_1, \cdots , m_d} \neq\emptyset$
	by construction. Thus $N_{c^{-k}}(\mathbb{F})\geq N_{c^{-k}}(\mathbb{F}_k)$. Moreover by construction,
	$N_{c^{-k}}(\mathbb{F}_k)=|F_k|$.
	Therefore
	$$\dimb(\mathbb{F})=\lim_{k\rightarrow\infty} \frac{\log N_{c^{-k}}(\mathbb{F})}{k\log c}
	= \lim_{k\rightarrow\infty} \frac{\log N_{c^{-k}}(\mathbb{F}_k)}{k\log c}=
	\lim_{k\rightarrow\infty} \frac{|F_k|}{k\log c} = \frac{|Y|}{\log c}.$$

	Since  box dimension coincides with  Hausdorff dimension on self similar continuous fractals, this ends the proof.
\end{proof}

\begin{proof}[Proof of Theorem \ref{th:4_2}]
Let $s>\dimzeta$, $\epsilon >0$, and $C=\sum_{n\in A}(n+1)^{-s}$. Consider the following
$(s/2+\epsilon)$-$\gamma$-supergale $d$, where $d(wn)=d(w)\frac{(n+1)^{2\epsilon}}{4C}$
for $n\in A$.
For each $f\in F_A$, there is an $m_0$ such that $f(m)^{2\epsilon}>8C$ for each $m\geq m_0$.
Therefore, if $|w|=m$, $d(wf(m))>2d(w)$ and $F_A\subseteq \strSS[d]$.

For the other direction, let $t>\dim_{\gamma}(F_A)$ and let $d$ be a $t$-gale such that
$F_A \subseteq S^{\infty}[d]$. Then the supremum over all $w\in A^*$ of
$\inf_{n\in A,n>|w|}d(wn)/d(w)$ is greater that $1$ (otherwise we can construct $f$ in $F_A-S^{\infty}[d]$).
Thus $\sum_{n\in A}(n+1)^{-2t} <\infty$.
\end{proof}

%% file: proof5.tex
\section{Appendix -- Zeta-Dimension and Algorithmic Information}
\begin{proof}[Proof for Theorem \ref{th:5_1}]
Assume the hypothesis.  For each $s \in [0,\infty )$,
$a \in \Delta$, and $0\leq t \in \Z$, let
\[ \beta_{s} = \sum_{w \in B} k^{-s\abs{w}} \]
and
\[ g(s,a,t) = \sum_{(w_{1},\cdots ,w_{t})\in B^{t}}
 \mathrm{num}_{k}(aw_{1} \cdots w_{t})^{-s} . \]
Also, for each $\vec{w}=(w_{1}, \cdots ,w_{t})\in B^{t}$, write
\[ l(\vec{w})=\sum^{t}_{i=1}\abs{w_{i}}. \]
Then, for all such $s$, $a$, and $t$, we have
\begin{eqnarray*}
  g(s,a,t) & \leq & \sum_{\vec{w}\in B^{t}} \mathrm{num}_{k}(a0^{l(\vec{w})})^{-s} \\
           & =    & \sum_{\vec{w}\in B^{t}} (a k^{l(\vec{w})})^{-s} \\
           & =    & a^{-s} \sum_{\vec{w}\in B^{t}} \prod_{i=1}^{t} k^{-s\abs{w_{i}}} \\
           & =    & a^{-s} \beta_{s}^{t}
\end{eqnarray*}
and
\begin{eqnarray*}
  g(s,a,t) & \geq & \sum_{\vec{w}\in B^{t}} \mathrm{num}_{k}(a(k-1)^{l(\vec{w})})^{-s} \\
           & \geq & \sum_{\vec{w}\in B^{t}} ((a+1) k^{l(\vec{w})})^{-s} \\
           & =    & (a+1)^{-s} \sum_{\vec{w}\in B^{t}} \prod^{t}_{i=1} k^{-s\abs{w_{i}}} \\
           & =    & (a+1)^{-s} \beta^{t}_{s} .
\end{eqnarray*}
That is, for all $s\in [0,\infty )$, $a\in\Delta$, and $0\leq t \in \Z$,
\begin{equation}\label{eq:5_1}
  (a+1)^{-s} \beta_{s}^{t} \leq g(s,a,t) \leq a^{-s} \beta_{s}^{t}.
\end{equation}
Since $B$ is an instantaneous code, we have
\begin{equation}\label{eq:5_2}
  \zeta_{A} (s) = \sum_{a\in\Delta} \sum^{\infty}_{t=0} g(s,a,t)
\end{equation}
for all $s\in [0,\infty )$.  Putting \eqref{eq:5_1} and \eqref{eq:5_2} together gives
\[
  \sum_{a\in\Delta} (a+1)^{-s} \sum_{t=0}^{\infty} \beta^{t}_{s}
  \leq \zeta_{A} (s) \leq \sum_{a \in \Delta} a^{-s} \sum^{\infty}_{t=0} \beta^{t}_{s} \]
for all $s \in [0,\infty)$.  By our choice of $s^{*}$, then,
\[ s > s^{*} \Rightarrow \beta_{s} < 1 \Rightarrow \zeta_{A} (s) < \infty \]
and
\[ s \leq s^{*} \Rightarrow \beta_{s} \geq 1 \Rightarrow \zeta_{A} (s) = \infty . \]
Thus $\zetadim (A) = s^{*}$.
\end{proof}

\begin{proof}[Proof of Corollary \ref{co:5_2}]
Apply Theorem \ref{th:5_1} with $\Delta = \Gamma - \{0\}$ and $B=\Gamma$.
\end{proof}

\begin{lemma}[Kraft's inequality]\label{lm:kraft_k}
Let $s>0$. Let $d$ be an $s$-supergale.
Then $\left|S^1[d]\cap\{0,1\}^k\right|\leq 2^{sk}d(\lambda)$
for all $k\in\N$.
\end{lemma}
\begin{proof}
Let $A=S^1[d]\cap\{0,1\}^k$.
Since $d$ is an $s$-supergale, we have
for every $w\in A$, $d(w)\geq 1$.
By the definition of supergale, we know that
\[\sum_{w\in\{0,1\}^k} d(w)\leq 2^{sk}d(\lambda).\]
Therefore
\begin{align*}
|A|\cdot 1&\leq \sum_{w\in A} d(w)\\
&\leq \sum_{w\in \{0,1\}^k} d(w)\\
&\leq 2^{sk}d(\lambda).
\end{align*}
\end{proof}

\begin{proof}[Proof of Theorem \ref{th:5_5}]
Let $s>0$. First, we show that for any $s$-supergale $d$,
\[\dim_\zeta( \bnum(S^1[d] \cap 1\{0,1\}^*) )\leq s.\]
Let $A= \bnum(S^1[d] \cap 1\{0,1\}^*)$. Let $\epsilon>0$.
\begin{align*}
\zeta_A(s+\epsilon)
&=\sum_{x\in A}\frac{1}{x^{s+\epsilon}}
\leq \sum_{w\in S^1[d] \cap 1\{0,1\}^*}\frac{1}{\bnum(w)^{s+\epsilon}}\\
&\leq s^{s+\epsilon}\sum_{w\in S^1[d] \cap 1\{0,1\}^*}\frac{1}{2^{(s+\epsilon)|w|}}\\
&\leq s^{s+\epsilon}\sum_{k=0}^\infty\sum_{\substack{w\in S^1[d]\\|w|=k} }\frac{1}{2^{(s+\epsilon)|w|}}\\
&=s^{s+\epsilon}\sum_{k=0}^\infty \left|S^1[d]\cap\{0,1\}^k\right|\frac{1}{2^{(s+\epsilon)k}}\\
&\leq^{\text{by Lemma \ref{lm:kraft_k}}} s^{s+\epsilon}\sum_{k=0}^\infty 2^{sk}d(\lambda)\frac{1}{2^{(s+\epsilon)k}}\\
&=s^{s+\epsilon}d(\lambda)\sum_{k=0}^\infty \frac{1}{2^{\epsilon k}}<\infty.
\end{align*}
Since $\epsilon$ is arbitrary, $\dim_\zeta(A)\leq s$.

Now we prove that if $\dim_\zeta(A)<s$,
then there exists an $s$-supergale $d$ such that
$A\subseteq \bnum(S^1[d])$.

Since $\dim_\zeta(A)<s$, for some $\epsilon>0$,
\[\sum_{k=0}^\infty  \frac{ \left| A_{=k}\right|}{2^{(s-\epsilon)k}} = 
\sum_{k=0}^\infty\sum_{\substack{w\in\{0,1\}^k\\ \bnum(w)\in A_{=k}} }\frac{1}{2^{(s-\epsilon)k}}\leq
\sum_{x\in A}\frac{1}{x^{s-\epsilon}}=\zeta_A(s-\epsilon)<\infty.\]
Thus there exists $n_0\in\Z^+$, such that for all $k>n_0$,
\[\frac{ \left| A_{=k}\right|}{2^{(s-\epsilon)k}} < 1.\]
Let
\[C_0=\max\left\{1, \frac{ \left| A_{=1}\right|}{2^{(s-\epsilon)1}},\frac{ \left| A_{=2}\right|}{2^{(s-\epsilon)2}},
\dots, \frac{ \left| A_{=n_0}\right|}{2^{(s-\epsilon)n_0}}  \right\}.\]
Let
\[C_1 = \max_{n\in\Z^+}\left\{\frac{n^2}{2^{\epsilon n}}\right\}.\]
Since $\frac{n^2}{2^{\epsilon n}}$ is eventually monotone decreasing,
$C_1<\infty$ exists.

We construct an $s$-supergale as follows.

For every $k\in\Z^+$, let $d_k:\{0,1\}^*\rightarrow [0,\infty)$ be defined by
the following recursion. And without loss of generality, for our convenience,
we assume that $|A_{=k}|\geq 1$ for all $k\in\Z^+$.
\[
d_k(w)=\begin{cases}
\frac{2^k}{|A_{=k}|},&|w|=k\text{ and }w\in A_{=k},\\
0,&|w|=k\text{ and }w\notin A_{=k},\\
\frac{d_k(w0)+d_k(w1)}{2},&|w|<k,\\
d_k(w[0..k-1]),&|w|>k.
\end{cases}
\]

Let
\[d(w)=C_0C_12^{(s-1)|w|}\sum_{k=1}^\infty\frac{1}{k^2}d_k(w).\]

It is easy to verify that $d_k$'s are martingales and
$d$ is an $s$-supergale.

Now let $x\in A$ and assume $x=\bnum(w)$ and $|w|=n\in \Z^+$.
\begin{align*}
d(w)&=C_0C_12^{(s-1)|w|}\sum_{k=0}^\infty\frac{1}{k^2}d_k(w)
\geq C_0C_12^{(s-1)n}\frac{1}{n^2}d_n(w)\\
&=C_0C_12^{(s-1)n}\frac{1}{n^2}\frac{2^n}{|A_{=n}|}
\geq C_0C_12^{(s-1)n}\frac{1}{n^2}\frac{2^n}{C_02^{(s-\epsilon)n}}\\
&=C_1\frac{2^{\epsilon n}}{n^2}\geq 1.
\end{align*}
Therefore, $w\in S^1[d]$, i.e., $x=\bnum(w) \in \bnum(S^1[d])$.
\end{proof}

\begin{theorem}\label{th:sum_pointwise}
Let $\alpha,\beta,\gamma \in[0,1]$ and
$\alpha < \beta \leq \gamma\leq \min\{1,\alpha+\beta\}$,
then there exist $A, B\subseteq \Z^+$ such that
$\dimzeta(A)=\alpha$, $\dimzeta(B)=\beta$
and $\dimzeta(A+B)=\gamma$.
\end{theorem}
\begin{proof}

Let
\[A_1=\{ x\in\Z^+ \mid x\geq 2^{|\rep_2(x)|-1}
\text{ and }x<2^{|\rep_2(x)|-1}+\ceil{2^{\alpha |\rep_2(x)|}}  \}\]
Let
\[B_1=
\{ x\in\Z^+ \mid x\geq 2^{|\rep_2(x)|-1}
\text{ and }x<2^{|\rep_2(x)|-1}+\ceil{2^{\beta |\rep_2(x)|}}  \}
\]
and
\[
B_2=
\{ x\in\Z^+\mid 
x=2^{|\rep_2(x)|-1}+k\floor{2^{\alpha|\rep_2(x)|}}, 0\leq 
k< \ceil{2^{(\gamma-\alpha)|\rep_2(x)|}} 
\}
\]

Let $T:\Z^+\rightarrow \Z^+$ be such that
$T(1)=1$ and $T(n+1)=2^{T(n)}$.

Let
\[B=(B_1\cup B_2)\cap\{ x\mid |x|=T(n)\text{ for some $n\in\Z^+$}\}\]
and
\[A=A_1\cap \{ x\mid |x|=T(n)\text{ for some $n\in\Z^+$}\}.\]

Let $C=A+B$. Let $n=T(k)$ for some $k\in\Z^+$.
Then
\[\myset{x}{ 
x\geq 2^{n-1}+2^{n-1}
\text{ and } x< 2^{n-1}+
\ceil{2^{\alpha n}}+
2^{n-1}+\floor{2^{\alpha n}}\ceil{2^{(\gamma-\alpha) n}-1}
}= C_{=n+1},\]
i.e.,
\[
\myset{ x}{
x\geq 2^n
\text{ and } x< 2^n
+\ceil{2^{\alpha n}}+
\floor{2^{\alpha n}}\ceil{2^{(\gamma-\alpha) n}-1}
 }=C_{=n+1},\]
and
\[C_{=n}\subseteq B_{=n} + A_{\leq\log n}.\]
It is easy to verify that
\[|C_{=n}|\leq |B_{=n} + A_{\leq\log n}|\leq 
n\ceil{2^{(\gamma-\alpha)n}}\]
and
\[
|C_{=n+1}| = \ceil{2^{\alpha n}}
+\floor{2^{\alpha n}}\ceil{2^{(\gamma-\alpha)n}-1}
,\]
i.e.,
\[2^{\gamma n}-2^{(\gamma-\alpha)n}\leq |C_{=n+1}|\leq 2\cdot 2^{\gamma n}.\]
For $n\neq T(k)$ and $n\neq T(k)+1$ for some $k\in\Z^+$,
it is easy to verify that $C_{=n}=\varnothing$.
It is now clear that the entropy rate of $C$
\[\eH_C=\limsup_{n\rightarrow \infty}
\frac{\log|C_{=n+1}|}{n+1}=
\limsup_{k\rightarrow\infty}\frac{\log|C_{=T(k)+1}|}{T(k)+1}=\gamma,
\]
i.e, $\dimzeta(C)=\gamma$.
Similarly, it is easy to verify that $\dimzeta(A)=\alpha$
and $\dimzeta(B)=\beta$.
\end{proof}

\begin{proof}[Proof of Theorem \ref{th:5_6}]
Let $\alpha=\dimzeta(A)$, $\beta=\dimzeta(B)$
and without loss of generality assume $\alpha\geq \beta$.
By Theorem \ref{th:sum_pointwise} and Staiger's proof that $\dimzeta(A*B)=\max\{\dimzeta(A), \dimzeta(B)\}$ \cite{Stai93},
it suffices to show that
\[\max\{\alpha, \beta\}\leq \dimzeta(A+B)\]
and
\[\dimzeta(A+B)\leq \alpha+\beta.\]

For the first inequality,
let $b=\min B$. Then it is easy to see that $\dimzeta(A+B)\geq \dimzeta(A+\{b\})$.
Since zeta-dimension is invariant under translation,
$\dimzeta(A+\{b\})=\dimzeta(A)=\alpha=\max\{\alpha, \beta\}$.

For the second inequality, let $\epsilon>0$.
Since $\dimzeta(A)=\alpha$ and $\dimzeta(B)=\beta$,
there exists $n_0\in\N$ such that
for all $n\geq n_0$,
$|A_{=n}|\leq 2^{(\alpha+\epsilon)n}$
and $|B_{=n}|\leq 2^{(\beta+\epsilon)n}$.
Let
\[C=\max\{\sum_{n=1}^{n_0-1}|A_{=n}|,\sum_{n=1}^{n_0-1}|B_{=n}|\}.\]
It is clear that
\begin{align*}
|(A+B)_{=n}|
&\leq (|A_{=n}|+|A_{=n-1}|)\sum_{k=1}^n|B_{=n}|
    + (|B_{=n}|+|B_{=n-1}|)\sum_{k=1}^n|A_{=n}|\\
&\leq (|A_{=n}|+|A_{=n-1}|)\sum_{k=n_0}^n|B_{=n}|
    + (|B_{=n}|+|B_{=n-1}|)\sum_{k=n_0}^n|A_{=n}|\\
    &\;\;\;\;+ C(|A_{=n}|+|A_{=n-1}|)
    + C(|B_{=n}|+|B_{=n-1}|)\\
&\leq
(1+2^{\alpha+\epsilon})2^{(\alpha+\epsilon)n}
\frac{2^{(\beta+\epsilon)n_0}(2^{(\beta+\epsilon)(n-n_0)}-1) }
{2^{\beta+\epsilon}}\\
&\;\;\;\;+(1+2^{\beta+\epsilon})2^{(\beta+\epsilon)n}
\frac{2^{(\alpha+\epsilon)n_0}(2^{(\alpha+\epsilon)(n-n_0)}-1) }
{2^{\alpha+\epsilon}}\\
&\;\;\;\;+C(1+2^{\alpha+\epsilon})2^{(\alpha+\epsilon)n}+
C(1+2^{\beta+\epsilon})2^{(\beta+\epsilon)n}.
\end{align*}
Let
\[C'=\max\{
C(1+2^{\alpha+\epsilon}2^{(\beta+\epsilon)n_0}),
C(1+2^{\beta+\epsilon}2^{(\alpha+\epsilon)n_0})
\}.\]
Then for all $n\geq n_0$
\[
|(A+B)_{=n}|
\leq C'2^{(\alpha+\beta+2\epsilon)n}.
\]
By the entropy characterization of zeta-dimension,
it is clear that $\dimzeta(A+B)\leq \alpha+\beta$.
\end{proof}